\documentclass[aps,showpacs,pra,superscriptaddress]{revtex4-1} 
\usepackage{bm}
\usepackage{color}
\usepackage{array}
\usepackage{multirow}
\usepackage{amsmath}
\usepackage{amssymb}
\usepackage{graphicx}
\usepackage{hyperref}

\makeatletter

\providecommand{\tabularnewline}{\\}

\makeatother

\begin{document}

\title{Self-to-self transitions in open quantum systems: the origin and solutions}

\author{Yaoxiong~Wang} \author{Ling~Yang}
\affiliation{Institute of Intelligent Machines, Chinese Academy of Sciences, Hefei 230031, China}
\affiliation{University of Science and Technology of China, Hefei 230027, China}

\author{Ying~Wang}
\affiliation{School of Mechanical Engineering, Shanghai Dianji University, Shanghai 201306, China}

\author{Shouzhi~Li} \author{Dewen~Cao}
\affiliation{Institute of Intelligent Machines, Chinese Academy of Sciences, Hefei 230031, China}
\affiliation{University of Science and Technology of China, Hefei 230027, China}

\author{Qing~Gao}
\affiliation{Institute of Intelligent Machines, Chinese Academy of Sciences, Hefei 230031, China}

\author{Feng~Shuang}
\affiliation{Institute of Intelligent Machines, Chinese Academy of Sciences, Hefei 230031, China}
\affiliation{University of Science and Technology of China, Hefei 230027, China}
\affiliation{Department of Mechanical Engineering, Anhui Polytechnic University, Wuhu 241000, China}

\author{Fang~Gao}
\email{gaofang@iim.ac.cn}
\affiliation{Institute of Intelligent Machines, Chinese Academy of Sciences, Hefei 230031, China}

\begin{abstract}
    The information of quantum pathways can be extracted in the framework of the Hamiltonian-encoding and Observable-decoding method.
    For closed quantum systems, only off-diagonal elements of the Hamiltonian in the Hilbert space is required to be encoded to obtain the desired transitions.
    For open quantum systems, environment-related terms will appear in the diagonal elements of the Hamiltonian in the Liouville space.
    Therefore, diagonal encodings have to be performed to differentiate different pathways, which will lead to self-to-self transitions and inconsistency of pathway amplitudes with Dyson expansion.
    In this work, a well-designed transformation is proposed to avoid the counter-intuitive transitions and the inconsistency, with or without control fields.
    A three-level open quantum system is employed for illustration, and numerical simulations show that the method are consistent with Dyson expansion.
\end{abstract}
\maketitle

\section{Introduction}
The control of quantum systems, particularly utilizing optimization techniques, has been more and more successful~\cite{key-1-1,key-1-2,key-1-3,key-1-4,key-1-5,key-1-6,key-1-7,key-1-8,key-1-9,key-1-10,key-1-11}.
Understanding the mechanisms that achieve the optimal control is becoming more interesting~\cite{key-1-2,m2,m3,m4,m5,m6,m7,key-1,key-2,key-3,PRA2014,key-1-14shuang-gao}.
Hamiltonian-encoding and Observable-decoding (HE-OD) is an efficient tool to real the underlying mechanism in the control of quantum systems~\cite{key-1-12}.
Rabitz et al. already demonstrate the application of HE-OD in the experiments~\cite{key-1-13,RobertoNJP}.
With this method, the mechanism is expressed with significant pathways linking the initial and target state.
In practice, the system Hamiltonian is encoded in a particular manner such that the resultant nonlinear distortion of the observable can be decoded to extract the pathway amplitudes.
Pathways in closed systems can be differentiated by performing only off-diagonal encodings, and their accurate amplitudes given by Dyson expansion are consistent with those by HE-OD.
Recently, we have expanded this method to open quantum systems to investigate the cooperation between an applied field and the environment~\cite{key-1-15gao}.
The dynamics of the open quantum systems is described in the Liouville space, and environment-related terms will mix with transition energies in some diagonal elements of the Hamiltonian.
If these diagonal elements are not encoded, pathways including the corresponding self-to-self transitions may not be differentiated.
This can be indicated by the inconsistency between the pathway amplitudes by HE-OD and Dyson expansion, as shown in subsection~\ref{subsec:Pathway-amplitudes-of}.
However, the self-to-self transitions are counter-intuitive and hard to understand.
To avoid this kind of transitions, a well-designed transformation is performed on the Hamiltonian to remove the diagonal elements.
The pathway amplitudes with this modified HE-OD method are almost the same as those with Dyson expansion.
Hence this improved method can be employed to differentiate pathways in open quantum systems with a clearer understanding of the control mechanism. 

This paper is arranged as follows.
Sec.~\ref{sec:The-HE-OD} describes the primitive HE-OD method with only diagonal encoding in open quantum systems and the inconsistency between the pathway amplitudes obtained by HE-OD and Dyson expansion.
In Sec.~\ref{sec:The-HE-OD-1}, diagonal encoding is included in HE-OD and the self-to-self transitions are introduced.
In Sec.~\ref{sec:The-research-on}, a well-designed transformation is adopted to deal with a three-level open quantum system with or without control fields, then the modified HE-OD method is generalized.
Sec.~\ref{sec:Conclusions-and-discussions} gives the conclusions. 

\section{Off-diagonal encoding in Open quantum systems\label{sec:The-HE-OD}}

\subsection{HE-OD in open systems \label{subsec:The-original-HE-OD}}

The Markovian dynamics of an open quantum system can be described by the Lindblad master equation 
\begin{equation}
    \frac{\partial\rho(t)}{\partial t}=-i[H_{0}-\mu E(t),\rho(t)]+\eta\mathcal{F}\{\rho(t)\},\label{eq:lindblad}
\end{equation}
Here $H_{0}$ is the unperturbed Hamiltonian, and $\mu$ and $E(t)$ are, respectively, the transition dipole operator and the control field.
For convenience $\hbar$ has been absorbed into $H_{0}$ and $\mu$, $\rho(t)$ is density operator and can be expressed with the eigenstates of $H_{0}$ as $\rho(t)=\underset{nm}{\sum}\rho_{nm}(t)\left|n\right\rangle \left\langle m\right|$.
The Lindblad term is
\begin{equation}
    \mathcal{F}\{\rho(t)\}= \sum_{j=1}^{d^{2}-1}
    (L_{j}\rho L_{j}^{+}-\frac{1}{2}L_{j}^{+}L_{j}\rho-\frac{1}{2}\rho L_{j}^{+}L_{j}),\label{eq:dissipation terms}
\end{equation}
where $\{L_{j}\}$ and $\eta$ are, respectively, Lindblad operators and the system-environmental coupling strength parameter.
It is convenient to rewrite the equation in the Liouville space as
\begin{equation}
    i\frac{\partial\rho_{jk}(t)}{\partial t}=\underset{m,n}{\sum}\mathcal{H}_{jk,mn}(t)\rho_{nm}(t).
    \label{eq:final lindblad}
\end{equation}
Here the density operator is expressed with the double-bracket notation,
$\left|\left.\rho(t)\right\rangle \right\rangle =\underset{nm}{\sum}\rho_{nm}(t)\left|\left.mn\right\rangle \right\rangle $.
The density operator at time $t$ can be derived from the non-unitary evolution operator $\mathcal{U}(t)$ and the density operator at time
$0$ as $\left|\left.\rho(t)\right\rangle \right\rangle =\mathcal{U}(t)\left|\left.\rho(0)\right\rangle \right\rangle $.
The master equation of $\mathcal{U}(t)$ is 
\begin{equation}
    i\frac{d\mathcal{U}(t)}{dt}=\mathcal{H}(t)\mathcal{U}(t),
    \label{eq:nonunitary equation in lindblad}
\end{equation}
Its solution can be expressed in the form of the Dyson expansion 
\begin{equation}
    \mathcal{U}(t) = I + (-i) \int_0^t \mathcal{H}(t_{1})dt_{1}
    +(-i)^{2} \int_0^t \mathcal{H}(t_{2}) \int_0^{t_2} \mathcal{H}(t_{1})dt_{1}dt_{2}+\cdots.
\end{equation}
Then the transition amplitude from the initial state $\left|\left.aa\right\rangle \right\rangle $ (\emph{i.e.}, 
state $\left|a\right\rangle $ in Hilbert space) to final state $\left|\left.bb\right\rangle \right\rangle $ (\emph{i.e.},
state $\left|b\right\rangle $ in Hilbert space) is
\begin{equation}
    \left\langle \left\langle bb\right.\right|\mathcal{U}(t)\left|\left.aa\right\rangle \right\rangle =\underset{n,\{l_{p}l_{q},\cdots\}}{\sum}\mathcal{U}_{bb,aa}^{n\{l_{n-1}^{'}l_{n-1},\cdots l_{1}^{'}l_{1}\}}(t),
    \label{eq:hilbert space}
\end{equation}
with 
\begin{align}
    & \mathcal{U}_{bb,aa}^{n\{l_{n-1}^{'}l_{n-1},\cdots l_{1}^{'}l_{1}\}}(t)
    =(-i)^{n} \int_0^t \left\langle \left\langle bb\right.\right|\mathcal{\mathcal{H}}(t_{n})\left|\left.l_{n-1}^{'}l_{n-1}\right\rangle \right\rangle \nonumber \\
    & \times \int_0^{t_n} \left\langle \left\langle l_{n-1}^{'}l_{n-1}\right.\right|\mathcal{\mathcal{H}}(t_{n-1})\left|\left.l_{n-2}^{'}l_{n-2}\right\rangle \right\rangle \times\cdots\times
    \int_0^{t_2} \left\langle \left\langle l_{1}^{'}l_{1}\right.\right|\mathcal{\mathcal{H}}(t_{1})\left|\left.aa\right\rangle \right\rangle dt_{1}\cdots dt_{n-1}dt_{n}.
\end{align}
This amplitude corresponds to a $n$th-order pathway from the initial state
$\left|\left.aa\right\rangle \right\rangle $ to final state
$\left|\left.bb\right\rangle \right\rangle $ through the set of $n$ intermediate steps
$\left|\left.aa\right\rangle \right\rangle \rightarrow\left|\left.l_{1}^{'}l_{1}\right\rangle \right\rangle \rightarrow\cdots\rightarrow\left|\left.l_{n-2}^{'}l_{n-2}\right\rangle \right\rangle \rightarrow\left|\left.l_{n-1}^{'}l_{n-1}\right\rangle \right\rangle \rightarrow\left|\left.bb\right\rangle \right\rangle $.
In HE-OD, the Hamiltonian is encoded as $\mathcal{H}_{ij}(t)\rightarrow\mathcal{H}_{ij}(t)m_{ij}(s)$,
where $m_{ij}(s)=exp(2\pi i\gamma_{ij}s/N)$ and $\gamma_{ij}$ is an element of the encoding matrix $\Gamma$.
After modulation, Eq.~\eqref{eq:nonunitary equation in lindblad} becomes
\begin{align}
    i\frac{d\mathcal{U}(t,s)}{dt} & =\left(\begin{array}{ccc}
        \mathcal{H}_{11}(t)m_{11}(s) & \cdots & \mathcal{H}_{11}(t)m_{1d}(s)\\
        \cdot & \cdot & \cdot\\
        \cdot & \cdot & \cdot\\
        \cdot & \cdot & \cdot\\
        \mathcal{H}_{d1}(t)m_{11}(s) & \cdots & \mathcal{H}_{dd}(t)m_{dd}(s)
    \end{array}\right)\mathcal{U}(t,s),
\end{align}
leading to the following transition amplitude
\begin{align}
    \mathcal{U}_{bb,aa}^{n\{l_{n-1}^{'}l_{n-1},\cdots l_{1}^{'}l_{1}\}}(t,s) 
    & = \int_0^T \cdots \int_0^{t_n}
    \mathcal{H}_{bb,l_{_{n-1}}^{'}l_{n-1}}(t)m_{bb,l_{_{n-1}}^{'}l_{n-1}}(s)\cdots\nonumber \\
    & \mathcal{\times H}_{l_{1}^{'}l_{1},aa}(t)m_{l_{1}^{'}l_{1},aa}(s)dt_{1}dt_{2}\cdots dt_{n}\nonumber \\
    & =\underset{n,\{l_{p}l_{q},\cdots\}}{\sum}\mathcal{U}_{bb,aa}^{n\{l_{n-1}^{'}l_{n-1},\cdots l_{1}^{'}l_{1}\}}(t)M{}_{bb,aa}^{n\{l_{n-1}^{'}l_{n-1},\cdots l_{1}^{'}l_{1}\}},
    \label{eq:u(t,s)}
\end{align}
Therefore, each particular pathway amplitude 
$\mathcal{U}_{bb,aa}^{n\{l_{n-1}^{'}l_{n-1},\cdots l_{1}^{'}l_{1}\}}(t)$
is labelled by its encoding function 
$M_{bb,aa}^{n\{l_{n-1}^{'}l_{n-1},\cdots l_{1}^{'}l_{1}\}}(s)=exp(2\pi i\gamma_{n(l_{n-1}^{'}l_{n-1},l_{n-2}^{'}l_{n-2},\cdots,l_{1}^{'}l_{1})})$.
Due to the orthogonal relationship between different encoding functions,
$\mathcal{U}_{bb,aa}^{n\{l_{n-1}^{'}l_{n-1},\cdots l_{1}^{'}l_{1}\}}(t)$
can be easily calculated by the inverse fast Fourier transform of
modulated matrix elements $\mathcal{U}_{bb,aa}^{n\{l_{n-1}^{'}l_{n-1},\cdots l_{1}^{'}l_{1}\}}(t,s)$
with the feature frequency of $\gamma_{n(l_{n-1}^{'}l_{n-1},l_{n-2}^{'}l_{n-2},\cdots,l_{1}^{'}l_{1})}=\gamma_{bb,l_{n-1}^{'}l_{n-1}}+\gamma_{l_{n-1}^{'}l_{n-1},l_{n-2}^{'}l_{n-2}}+\cdots+\gamma_{l_{1}^{'}l_{1},aa}$.

\subsection{Off-diagonal encoding scheme and the inconsistency\label{subsec:Pathway-amplitudes-of}}

\begin{figure}
    \includegraphics[width=0.3\columnwidth]{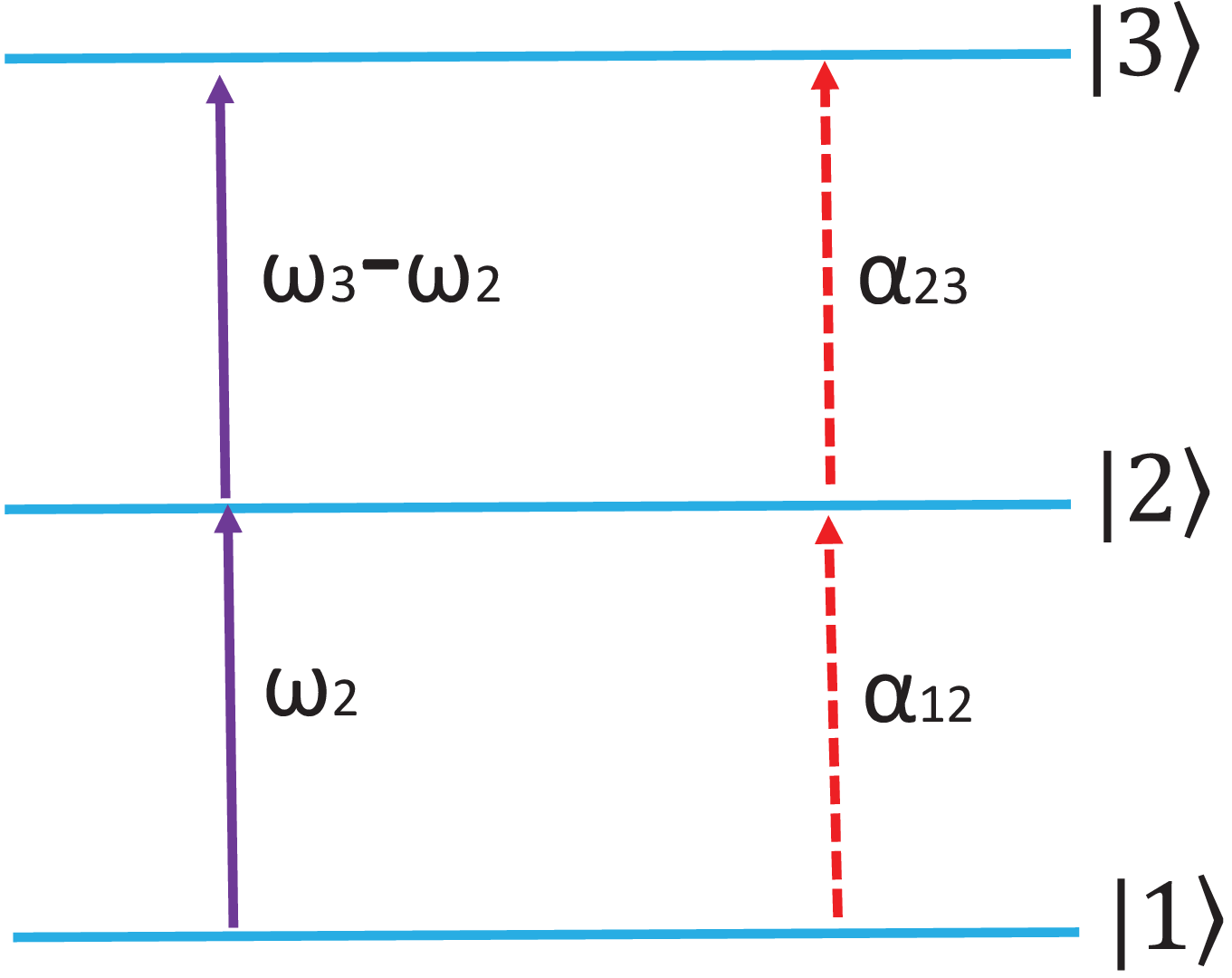}
    \caption{\label{fig:the-structure-of-open-system}
    (Color online) A three-level open model system.
The purple solid arrows represent dipole-induced transitions, while the red dashed arrows correspond to dissipation-induced transitions.  }
\end{figure}

In this subsection, a three-level open system is taken as example to demonstrate the HE-OD, where only off-diagonal terms are encoded.
The inconsistency between the pathway amplitudes by HE-OD and Dyson expansion is shown.
The model system is in Fig.~\ref{fig:the-structure-of-open-system}, and the corresponding matrices in Eq.~\eqref{eq:lindblad} are 
\begin{equation}
    H_{0}=\left(\begin{array}{ccc}
        0 & 0 & 0\\
        0 & \omega_{2} & 0\\
        0 & 0 & \omega_{3}
    \end{array}\right),\mu=\left(\begin{array}{ccc}
        0 & \mu_{12} & 0\\
        \mu_{12} & 0 & \mu_{23}\\
        0 & \mu_{23} & 0
    \end{array}\right),
\end{equation}
\begin{equation}
    L_{1}=L_{2}^{+}=\left(\begin{array}{ccc}
        0 & \sqrt{\alpha_{12}} & 0\\
        0 & 0 & 0\\
        0 & 0 & 0
    \end{array}\right),
\end{equation}
\begin{equation}
    L_{3}=L_{4}^{+}=\left(\begin{array}{ccc}
        0 & 0 & 0\\
        0 & 0 & \sqrt{\alpha_{23}}\\
        0 & 0 & 0
    \end{array}\right),
\end{equation}
It can be seen that only the nearest-neighbor transitions are induced by the dipole or Lindblad operators.
Then the Hamiltonian matrix $\mathcal{H}(t)$ in the Liouville space is
\begin{equation}
    \mathcal{H}(t)=\left(\begin{array}{ccccccccc}
        \beta_{1} & \chi_{12} & 0 & -\chi_{12} & i\eta\alpha_{12} & 0 & 0 & 0 & 0\\
        \chi_{12} & \beta_{2} & \chi_{23} & 0 & -\chi_{12} & 0 & 0 & 0 & 0\\
        0 & \chi_{23} & \beta_{3} & 0 & 0 & -\chi_{12} & 0 & 0 & 0\\
        -\chi_{12} & 0 & 0 & \beta_{4} & \chi_{12} & 0 & -\chi_{23} & 0 & 0\\
        i\eta\alpha_{12} & -\chi_{12} & 0 & \chi_{12} & \beta_{5} & \chi_{23} & 0 & -\chi_{23} & i\eta\alpha_{23}\\
        0 & 0 & -\chi_{12} & 0 & \chi_{23} & \beta_{6} & 0 & 0 & -\chi_{23}\\
        0 & 0 & 0 & -\chi_{23} & 0 & 0 & \beta_{7} & \chi_{12} & 0\\
        0 & 0 & 0 & 0 & -\chi_{23} & 0 & \chi_{12} & \beta_{8} & \chi_{23}\\
        0 & 0 & 0 & 0 & i\eta\alpha_{23} & -\chi_{23} & 0 & \chi_{23} & \beta_{9}
    \end{array}\right),\label{eq:hamiltonian matrix}
\end{equation}
with $\beta_{1}=-i\eta\alpha_{12}$, $\beta_{2}=-\omega_{2}-i\eta(\alpha_{12}+\frac{\alpha_{23}}{2})$,
$\beta_{3}=-\omega_{3}-\frac{i}{2}\eta(\alpha_{12}+\alpha_{23})$,
$\beta_{4}=\omega_{2}-i\eta(\alpha_{12}+\frac{\alpha_{23}}{2})$,
$\beta_{5}=-i\eta(\alpha_{12}+\alpha_{23})$, $\beta_{6}=\omega_{2}-\omega_{3}-i\eta(\frac{\alpha_{12}}{2}+\alpha_{23})$,
$\beta_{7}=\omega_{3}-\frac{i}{2}\eta(\alpha_{12}+\alpha_{23})$,
$\beta_{8}=\omega_{3}-\omega_{2}-i\eta(\frac{\alpha_{12}}{2}+\alpha_{23})$,
$\beta_{9}=-i\eta\alpha_{23}$, and $\chi_{ij}=\mu_{ij}E(t)$. The
control field is taken to be a Gaussian form as $E(t)=e^{-\frac{(t-T/2)^{2}}{2\sigma^{2}}}\underset{l}{\sum}A_{l}cos(\upsilon_{l}t+\theta_{l})$,
where $\upsilon_{1}=\omega_{2}$ and $\upsilon_{2}=\omega_{3}-\omega_{2}$ are resonant frequencies of the two allowed transitions.
The parameters are $T=8268.221(200\mathrm{f}s)$ , $\sigma=1240.23(30\mathrm{f}s)$,
$\omega_{2}=0.0365$, $\omega_{3}=0.0651$, $\mu_{12}=0.0691$, $\mu_{23}=0.0835$,
$A_{1}=0.0038$, $A_{2}=0.0037$, $\theta_{1}=1.6551$, $\theta_{2}=3.2031$,
$\alpha_{12}=0.089$ and $\alpha_{23}=0.194$. The encoding matrix is taken to be 
\begin{equation}
    \Gamma=\left(\begin{array}{ccccccccc}
        0 & 1 & 0 & 5 & 17 & 0 & 0 & 0 & 0\\
        1 & 0 & 21 & 0 & 33 & 0 & 0 & 0 & 0\\
        0 & 21 & 0 & 0 & 0 & 41 & 0 & 0 & 0\\
        5 & 0 & 0 & 0 & 59 & 0 & 68 & 0 & 0\\
        17 & 33 & 0 & 59 & 0 & 77 & 0 & 83 & 101\\
        0 & 0 & 41 & 0 & 77 & 0 & 0 & 0 & 109\\
        0 & 0 & 0 & 68 & 0 & 0 & 0 & 111 & 0\\
        0 & 0 & 0 & 0 & 83 & 0 & 111 & 0 & 127\\
        0 & 0 & 0 & 0 & 101 & 109 & 0 & 127 & 0
    \end{array}\right).
\end{equation}
Here the diagonal elements of the Hamiltonian are not encoded. 
The significant pathways extracted by HE-OD are shown in Tab.~\ref{tab:the results of the HE-OD and Integration}.
There are three types of pathways: dipole-induced, dipole-environment-induced and environment-induced ones.
For example, the dipole-induced pathway$\left.\left|11\right\rangle \right\rangle \rightarrow\left.\left|21\right\rangle \right\rangle \rightarrow\left.\left|31\right\rangle \right\rangle \rightarrow\left.\left|32\right\rangle \right\rangle \rightarrow\left.\left|33\right\rangle \right\rangle $ is induced by two dipole operators ($\mu_{12}$ and $\mu_{23}$) according to Eq.~\eqref{eq:hamiltonian matrix},
and the dipole-environment-induced pathway $\left.\left|11\right\rangle \right\rangle \rightarrow\left.\left|21\right\rangle \right\rangle \rightarrow\left.\left|22\right\rangle \right\rangle \rightarrow\left.\left|33\right\rangle \right\rangle $
by one dipole operator $\mu_{12}$ and the environment-related term
$\alpha_{23}$. The environment-induced pathway pathway $\left.\left|11\right\rangle \right\rangle \rightarrow\left.\left|22\right\rangle \right\rangle \rightarrow\left.\left|33\right\rangle \right\rangle $
is related to the two environmental terms $\alpha_{12}$ and $\alpha_{23}$.
Some symmetry relation (\emph{i.e.} the same magnitude and opposite phase) can be found for pathways induced 
by the same operators (\emph{e.g.} pathways featured by the inverse fast Fourier transform (IFFT) frequencies of 244 and 250
in Tab.~\ref{tab:the results of the HE-OD and Integration}).
The results are shown in Tab.~\ref{tab:the results of the HE-OD and Integration} labelled with ''Off-diagonal Encoding Scheme''.  
\begin{table}
    \caption{\label{tab:the results of the HE-OD and Integration}Magnitudes and
    phases of significant quantum pathways. LF stands for the feature
    frequencies of the pathways. The values in the parentheses are the
phases. }

\begin{tabular}{ccc>{\centering}p{3cm}>{\centering}p{3cm}c}
    \hline 
    Type & Pathways & LF & Off-diagonal Encoding Scheme &  Scheme including Diagonal Encoding & Integration\tabularnewline
    \hline 
    \multirow{6}{*}{dipole-induced} & $\left.\left|11\right\rangle \right\rangle \rightarrow\left.\left|12\right\rangle \right\rangle \rightarrow\left.\left|13\right\rangle \right\rangle \rightarrow\left.\left|23\right\rangle \right\rangle \rightarrow\left.\left|33\right\rangle \right\rangle $ & 172 & $1.197\times10^{-3}$

    ($1.704\times10^{-2}$) & $1.433\times10^{-3}$

    ($1.606\times10^{-2}$) & $1.606\times10^{-3}$\tabularnewline
    & $\left.\left|11\right\rangle \right\rangle \rightarrow\left.\left|21\right\rangle \right\rangle \rightarrow\left.\left|31\right\rangle \right\rangle \rightarrow\left.\left|32\right\rangle \right\rangle \rightarrow\left.\left|33\right\rangle \right\rangle $ & 311 & $1.200\times10^{-3}$

    ($-1.714\times10^{-2}$) & $1.436\times10^{-3}$

    ($-1.616\times10^{-2}$) & $1.606\times10^{-3}$\tabularnewline
    & $\left.\left|11\right\rangle \right\rangle \rightarrow\left.\left|12\right\rangle \right\rangle \rightarrow\left.\left|22\right\rangle \right\rangle \rightarrow\left.\left|32\right\rangle \right\rangle \rightarrow\left.\left|33\right\rangle \right\rangle $ & 244 & $1.192\times10^{-3}$

    ($6.901\times10^{-2}$) & $1.492\times10^{-3}$

    ($7.884\times10^{-2}$) & $1.642\times10^{-3}$\tabularnewline
    & $\left.\left|11\right\rangle \right\rangle \rightarrow\left.\left|21\right\rangle \right\rangle \rightarrow\left.\left|22\right\rangle \right\rangle \rightarrow\left.\left|23\right\rangle \right\rangle \rightarrow\left.\left|33\right\rangle \right\rangle $ & 250 & $1.197\times10^{-3}$

    ($-6.816\times10^{-2}$) & $1.500\times10^{-3}$

    ($-7.778\times10^{-2}$) & $1.642\times10^{-3}$\tabularnewline
    & $\left.\left|11\right\rangle \right\rangle \rightarrow\left.\left|12\right\rangle \right\rangle \rightarrow\left.\left|22\right\rangle \right\rangle \rightarrow\left.\left|23\right\rangle \right\rangle \rightarrow\left.\left|33\right\rangle \right\rangle $ & 220 & $1.171\times10^{-3}$

    ($2.433\times10^{-2}$) & $1.468\times10^{-3}$

    ($2.476\times10^{-2}$) & $1.615\times10^{-3}$\tabularnewline
    & $\left.\left|11\right\rangle \right\rangle \rightarrow\left.\left|21\right\rangle \right\rangle \rightarrow\left.\left|22\right\rangle \right\rangle \rightarrow\left.\left|32\right\rangle \right\rangle \rightarrow\left.\left|33\right\rangle \right\rangle $ & 274 & $1.172\times10^{-3}$

    ($-2.422\times10^{-2}$) & $1.471\times10^{-3}$

    ($-2.465\times10^{-2}$) & $1.615\times10^{-3}$\tabularnewline
    \hline 
    \multirow{4}{*}{ Dipole-dissipation-induced} & $\left.\left|11\right\rangle \right\rangle \rightarrow\left.\left|12\right\rangle \right\rangle \rightarrow\left.\left|22\right\rangle \right\rangle \rightarrow\left.\left|33\right\rangle \right\rangle $ & 135 & $8.978\times10^{-3}$

    ($8.187\times10^{-2}$) & $1.166\times10^{-2}$

    ($8.536\times10^{-2}$) & $1.230\times10^{-2}$\tabularnewline
    & $\left.\left|11\right\rangle \right\rangle \rightarrow\left.\left|21\right\rangle \right\rangle \rightarrow\left.\left|22\right\rangle \right\rangle \rightarrow\left.\left|33\right\rangle \right\rangle $ & 165 & $8.978\times10^{-3}$

    ($-8.187\times10^{-2}$) & $1.166\times10^{-2}$

    ($-8.536\times10^{-2}$) & $1.230\times10^{-2}$\tabularnewline
    & $\left.\left|11\right\rangle \right\rangle \rightarrow\left.\left|22\right\rangle \right\rangle \rightarrow\left.\left|23\right\rangle \right\rangle \rightarrow\left.\left|33\right\rangle \right\rangle $ & 203 & $5.436\times10^{-3}$

    ($-5.526\times10^{-2}$) & $7.296\times10^{-3}$

    ($-5.811\times10^{-2}$) & $7.798\times10^{-3}$\tabularnewline
    & $\left.\left|11\right\rangle \right\rangle \rightarrow\left.\left|22\right\rangle \right\rangle \rightarrow\left.\left|32\right\rangle \right\rangle \rightarrow\left.\left|33\right\rangle \right\rangle $ & 227 & $5.435\times10^{-3}$

    ($5.530\times10^{-2}$) & $7.295\times10^{-3}$

    ($5.815\times10^{-2}$) & $7.798\times10^{-3}$\tabularnewline
    \hline 
    \multirow{1}{*}{Dissipation-induced} & $\left.\left|11\right\rangle \right\rangle \rightarrow\left.\left|22\right\rangle \right\rangle \rightarrow\left.\left|33\right\rangle \right\rangle $ & 118 & $2.032\times10^{-2}$

    ($6.794\times10^{-15}$) & $2.857\times10^{-2}$

    ($6.572\times10^{-15}$) & $2.858\times10^{-2}$\tabularnewline
    \hline 
\end{tabular}
\end{table}

\begin{figure}
    \includegraphics[width=0.8\columnwidth]{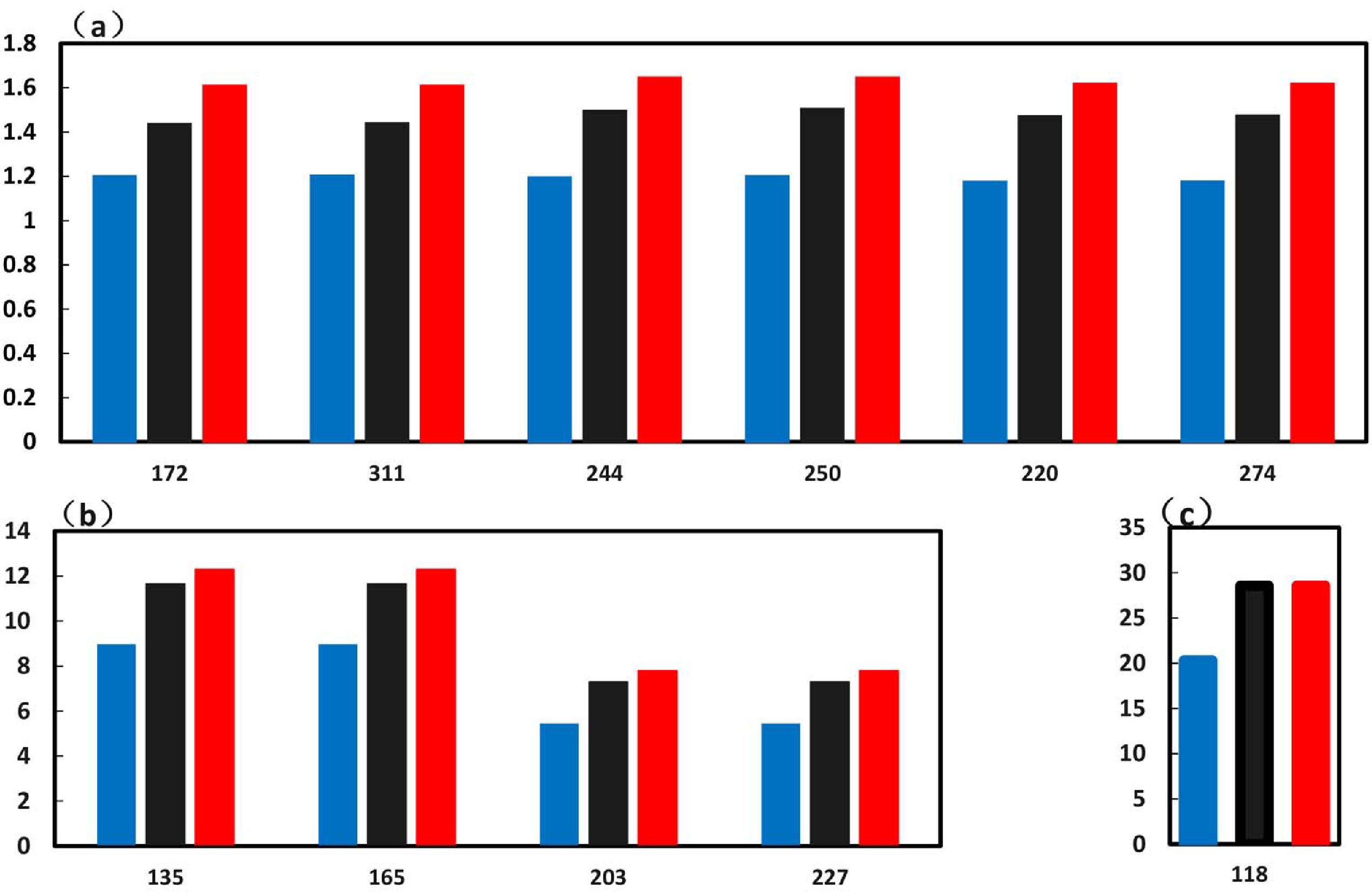}
    \caption{\label{fig:the-three-level-system-HE-OD}
    (Color online) Magnitudes of significant pathways with different methods.
    The vertical axis denotes the magnitude with the values scaled by a factor of 1000,
    while the horizontal axis labels the IFFT frequencies of the pathways.
    Panels (a), (b) and (c) are, respectively, the dipole-induced pathways,
    the dipole-environment-induced pathways and the exclusively environment-induced pathways. 
    The blue, black and red bars are, respectively, 
    obtained by off-diagonal encoding scheme, scheme including diagonal encoding and Dyson expansion. }
\end{figure}

In the interaction picture, the Hamiltonian $\mathcal{H}(t)$ is split to $\mathcal{H}_{0}+V(t)$,
then the master equation becomes 
\begin{equation}
    i\frac{d\mathcal{U}_{I}(t)}{dt}=V_{I}(t)\mathcal{U}_{I}(t),
    \label{eq:interaction Lindblad equation.}
\end{equation}
with $V_{I}(t)=e^{iH_{0}t}V(t)e^{-iH_{0}t}$ and $\mathcal{U}{}_{I}(t)=e^{iH_{0}t}\mathcal{U}(t)e^{-iH_{0}t}$. 

The matrices are
\begin{equation}
    \mathcal{H}_{0}=\left(\begin{array}{ccccccccc}
        0 & 0 & 0 & 0 & 0 & 0 & 0 & 0 & 0\\
        0 & -\omega_{2} & 0 & 0 & 0 & 0 & 0 & 0 & 0\\
        0 & 0 & -\omega_{3} & 0 & 0 & 0 & 0 & 0 & 0\\
        0 & 0 & 0 & \omega_{2} & 0 & 0 & 0 & 0 & 0\\
        0 & 0 & 0 & 0 & 0 & 0 & 0 & 0 & 0\\
        0 & 0 & 0 & 0 & 0 & \omega_{2}-\omega_{3} & 0 & 0 & 0\\
        0 & 0 & 0 & 0 & 0 & 0 & \omega_{3} & 0 & 0\\
        0 & 0 & 0 & 0 & 0 & 0 & 0 & \omega_{3}-\omega_{2} & 0\\
        0 & 0 & 0 & 0 & 0 & 0 & 0 & 0 & 0
    \end{array}\right),\label{eq:hamiltonian matrix-1}
\end{equation}
and
\begin{equation}
    V(t)=\left(\begin{array}{ccccccccc}
        \xi_{1} & \chi_{12} & 0 & -\chi_{12} & i\eta\alpha_{12} & 0 & 0 & 0 & 0\\
        \chi_{12} & \xi_{2} & \chi_{23} & 0 & -\chi_{12} & 0 & 0 & 0 & 0\\
        0 & \chi_{23} & \xi_{3} & 0 & 0 & -\chi_{12} & 0 & 0 & 0\\
        -\chi_{12} & 0 & 0 & \xi_{4} & \chi_{12} & 0 & -\chi_{23} & 0 & 0\\
        i\eta\alpha_{12} & -\chi_{12} & 0 & \chi_{12} & \xi_{5} & \chi_{23} & 0 & -\chi_{23} & i\eta\alpha_{23}\\
        0 & 0 & -\chi_{12} & 0 & \chi_{23} & \xi_{6} & 0 & 0 & -\chi_{23}\\
        0 & 0 & 0 & -\chi_{23} & 0 & 0 & \xi_{7} & \chi_{12} & 0\\
        0 & 0 & 0 & 0 & -\chi_{23} & 0 & \chi_{12} & \xi_{8} & \chi_{23}\\
        0 & 0 & 0 & 0 & i\eta\alpha_{23} & -\chi_{23} & 0 & \chi_{23} & \xi_{9}
    \end{array}\right),\label{eq:hamiltonian matrix-2}
\end{equation}
Here $\xi_{1}=\beta_{1}=-i\eta\alpha_{12}$, $\xi_{2}=-i\eta(\alpha_{12}+\frac{\alpha_{23}}{2})$,
$\xi_{3}=-\frac{i}{2}\eta(\alpha_{12}+\alpha_{23})$, $\xi_{4}=-i\eta(\alpha_{12}+\frac{\alpha_{23}}{2})$,
$\xi_{5}=\beta_{5}=-i\eta(\alpha_{12}+\alpha_{23})$, $\xi_{6}=-i\eta(\frac{\alpha_{12}}{2}+\alpha_{23})$,
$\xi_{7}=-\frac{i}{2}\eta(\alpha_{12}+\alpha_{23})$, $\xi_{8}=-i\eta(\frac{\alpha_{12}}{2}+\alpha_{23})$
and $\xi_{9}=\beta_{9}=-i\eta\alpha_{23}$.
Thus the Hamiltonian matrix in the interaction picture $V_{I}(t)$ can be derived as 
\begin{equation}
    V_{I}(t)=\left[\begin{array}{ccccccccc}
            \xi_{1} & \ell(t) & 0 & -\ell(-t) & i\eta\alpha_{12} & 0 & 0 & 0 & 0\\
            \ell(-t) & \xi_{2} & \kappa(-t) & 0 & -\ell(-t) & 0 & 0 & 0 & 0\\
            0 & \kappa(t) & \xi_{3} & 0 & 0 & -\ell(-t) & 0 & 0 & 0\\
            -\ell(t) & 0 & 0 & \xi_{4} & \ell(t) & 0 & -\kappa(t) & 0 & 0\\
            i\eta\alpha_{12} & -\ell(t) & 0 & \ell(-t) & \xi_{5} & \kappa(-t) & 0 & -\kappa(t) & i\eta\alpha_{23}\\
            0 & 0 & -\ell(t) & 0 & \kappa(t) & \xi_{6} & 0 & 0 & -\kappa(t)\\
            0 & 0 & 0 & -\kappa(-t) & 0 & 0 & \xi_{7} & \ell(t) & 0\\
            0 & 0 & 0 & 0 & -\kappa(-t) & 0 & \ell(-t) & \xi_{8} & \kappa(-t)\\
            0 & 0 & 0 & 0 & i\eta\alpha_{23} & -\kappa(-t) & 0 & \kappa(t) & \xi_{9}
    \end{array}\right],\label{eq:hamiltonian matrix-2-1}
\end{equation}
with $\kappa(t)=\chi_{23}e^{i(\omega_{2}-\omega_{3})t}$ and $\ell(t)=\chi_{12}e^{i\omega_{2}t}$.
The pathway amplitudes can also be obtained as the corresponding Dyson terms.
For example, with $U_{i}$ standing for the amplitude of pathway $i$ (\emph{i.e.} LF 
in Tab.~\ref{tab:the results of the HE-OD and Integration}),
we have 
\begin{align}
    U_{172}=
    & \int_0^T (-\chi_{23})e^{-i(\omega_{2}-\omega_{3})t_{4}}
    \int_0^{t_4} (-\chi_{12})e^{i\omega_{2}t_{3}}
    \int_0^{t_3} \chi_{23}e^{i(\omega_{2}-\omega_{3})t_{2}}
    \int_0^{t_2} \chi_{12}e^{-i\omega_{2}t_{1}}dt_{1}dt_{2}dt_{3}dt_{4},
\end{align}
\begin{equation}
    U_{135}=
    \eta\mu_{12}^{2}\alpha_{23}
    \int_0^T \int_0^{t_3} E(t_{2})e^{i\omega_{2}t_{2}}
    \int_0^{t_2} E(t_{1})e^{-i\omega_{2}t_{1}}dt_{1}dt_{2}dt_{3},
\end{equation}
\begin{equation}
    U_{118}=(-i)^{2} \int_0^T \eta\alpha_{12}
    \int_0^{t_2} \eta\alpha_{23}dt_{1}dt_{2}.
\end{equation}

The results are listed in Tab.~\ref{tab:the results of the HE-OD and Integration} labelled with ``Integration''.
There are non-negligible differences between results of ``Off-diagonal Encoding Scheme'' ($U_{H_{1}}$) and ``Integration'' ($U_{I}$).
The difference ratio $R=\frac{\left|U_{I}|-|U_{H_{1}}\right|}{\left|U_{I}\right|}$ can be as large as 30\%,
which is easier to see in Fig.~\ref{fig:the-three-level-system-HE-OD}. 

\section{the Origin of Self-to-self transitions\label{sec:The-HE-OD-1}}

In subsection \ref{subsec:The-original-HE-OD}, 
scheme with only off-diagonal encoding is adopted in HE-OD to calculate pathway amplitudes.
However, it is noticed that some diagonal elements of the Hamiltonian also contain environment-related terms (\emph{i.e.} $\alpha_{12}$ and $\alpha_{23}$).
Self-to-self transitions, such as $\left.\left|11\right\rangle \right\rangle \rightarrow\left.\left|11\right\rangle \right\rangle ,\left.\left|22\right\rangle \right\rangle \rightarrow\left.\left|22\right\rangle \right\rangle ,\left|33\right\rangle \rightarrow\left.\left|33\right\rangle \right\rangle $ are neglected improperly.
Thus diagonal encodings may be possible to diminish the difference between pathway amplitudes by the off-diagonal encoding scheme and Dyson expansion.
In our scheme including diagonal encoding, only the elements of $\mathcal{H_{\text{11}}}(t)$, $\mathcal{H_{\text{55}}}(t)$ and $\mathcal{H_{\text{99}}}(t)$ with pure dissipation terms are encoded, and the encoding matrix is
\begin{equation}
    \Gamma_{1}=\left(\begin{array}{ccccccccc}
        221 & 1 & 0 & 5 & 17 & 0 & 0 & 0 & 0\\
        1 & 0 & 21 & 0 & 33 & 0 & 0 & 0 & 0\\
        0 & 21 & 0 & 0 & 0 & 41 & 0 & 0 & 0\\
        5 & 0 & 0 & 0 & 59 & 0 & 68 & 0 & 0\\
        17 & 33 & 0 & 59 & 277 & 77 & 0 & 83 & 101\\
        0 & 0 & 41 & 0 & 77 & 0 & 0 & 0 & 109\\
        0 & 0 & 0 & 68 & 0 & 0 & 0 & 111 & 0\\
        0 & 0 & 0 & 0 & 83 & 0 & 111 & 0 & 127\\
        0 & 0 & 0 & 0 & 101 & 109 & 0 & 127 & 341
    \end{array}\right).\label{eq:first encoding matrix}
\end{equation}
The results are listed in Tab.~\ref{tab:the results of the HE-OD and Integration}
labelled with ``Scheme including Diagonal Encoding''.
As shown in Fig.~\ref{fig:the-three-level-system-HE-OD},
the difference between pathway amplitudes by this new encoding scheme and integration becomes smaller,
and the ratio $R=\frac{\left|U_{I}\right|-\left|U_{H_{2}}\right|}{\left|U_{I}\right|}$ is now about 10\%,
with $\left|U_{H_{2}}\right|$ stands for the magnitude by the scheme including diagonal encoding.
The smallest difference is only $\sim0.03\%$ for pathway $\left.\left|11\right\rangle \right\rangle \rightarrow\left.\left|22\right\rangle \right\rangle \rightarrow\left.\left|33\right\rangle \right\rangle$. 

As shown in Tab.~\ref{tab:the results of the HE-OD and Integration},
the new scheme will lead to significant pathways involving self-to-self transitions.
The discrepancy between the pathway amplitudes $U_{H_{1}}$ by the off-diagonal encoding scheme and $U_{I}$ 
by Dyson expansion in Tab.~\ref{tab:the results of the HE-OD and Integration} may come from these self-to-self transitions.
The off-diagonal encoding scheme,
which doesn't include diagonal encodings,
can not distinguish pathways with and without self-to-self transitions,
while the value of $U_{H_{1}}$ should be the sum amplitude of pathways without and with self-to-self transitions.
Therefore, the difference between $U_{H_{2}}$ by the scheme including diagonal encoding,
which considers this kind of transitions, and $U_{I}$ becomes smaller.
However, it is also noticed in Eq.~\eqref{eq:hamiltonian matrix}
that environmental terms are ``entangled'' with the resonant frequencies (\emph{i.e.} $\omega_{2}$, $\omega_{3}$) in the other diagonal Hamiltonian elements except $\mathcal{H_{\text{11}}}(t)$, $\mathcal{H_{\text{55}}}(t)$ and $\mathcal{H_{\text{99}}}(t)$.
The scheme including diagonal encoding does not consider this, which leads to a small discrepancy between $U_{H_{2}}$ and $U_{I}$.
It is necessary to disentangle the two types of terms and then encode all environment-related ones to further improve the accuracy. 

\section{Solutions and Demonstration \label{sec:The-research-on}}

It has been shown above that a proper encoding scheme has to be adopted to extract correct pathway amplitudes with HE-OD. However, this requires:
1. the environmental terms to be disentangled from resonant frequencies;
2. diagonal encodings to be performed, which results in self-to-self transitions.
Techniquely, the original definition of quantum pathways is jumping from one state to another state,
but the self-to-self transitions is quite counter-intuitive and not proper to be involved in the pathway,
because it is jumping from a state to itself. Eq.~\eqref{eq:hamiltonian matrix-2-1} indicates that moving to the interaction picture is a way to achieve the disentanglement, but still can not avoid diagonal encodings.
To conquer this problem, a mathematical transformation is applied to the Hamiltonian in the interaction picture firstly and then normal off-diagonal encoding schemes can be adopted.
The same three-level open model system in Fig.~\ref{fig:the-structure-of-open-system} with and without control fields is employed for illustration in this section.

\subsection{The three-level open quantum system without control field}

When there is no control field, the Hamiltonian in the Liouville space
becomes much simpler, 
\begin{equation}
    \mathcal{H}=\left(\begin{array}{ccccccccc}
        \beta_{1} & 0 & 0 & 0 & i\eta\alpha_{12} & 0 & 0 & 0 & 0\\
        0 & \beta_{2} & 0 & 0 & 0 & 0 & 0 & 0 & 0\\
        0 & 0 & \beta_{3} & 0 & 0 & 0 & 0 & 0 & 0\\
        0 & 0 & 0 & \beta_{4} & 0 & 0 & 0 & 0 & 0\\
        i\eta\alpha_{12} & 0 & 0 & 0 & \beta_{5} & 0 & 0 & 0 & i\eta\alpha_{23}\\
        0 & 0 & 0 & 0 & 0 & \beta_{6} & 0 & 0 & 0\\
        0 & 0 & 0 & 0 & 0 & 0 & \beta_{7} & 0 & 0\\
        0 & 0 & 0 & 0 & 0 & 0 & 0 & \beta_{8} & 0\\
        0 & 0 & 0 & 0 & i\eta\alpha_{23} & 0 & 0 & 0 & \beta_{9}
    \end{array}\right).\label{eq:three-level-hamiltonian-1-1}
\end{equation}

The resulted Hamiltonian in the interaction picture is 
\begin{equation}
    \mathcal{H}_{I}=\left(\begin{array}{ccccccccc}
        \xi_{1} & 0 & 0 & 0 & i\eta\alpha_{12} & 0 & 0 & 0 & 0\\
        0 & \xi_{2} & 0 & 0 & 0 & 0 & 0 & 0 & 0\\
        0 & 0 & \xi_{3} & 0 & 0 & 0 & 0 & 0 & 0\\
        0 & 0 & 0 & \xi_{4} & 0 & 0 & 0 & 0 & 0\\
        i\eta\alpha_{12} & 0 & 0 & 0 & \xi_{5} & 0 & 0 & 0 & i\eta\alpha_{23}\\
        0 & 0 & 0 & 0 & 0 & \xi_{6} & 0 & 0 & 0\\
        0 & 0 & 0 & 0 & 0 & 0 & \xi_{7} & 0 & 0\\
        0 & 0 & 0 & 0 & 0 & 0 & 0 & \xi_{8} & 0\\
        0 & 0 & 0 & 0 & i\eta\alpha_{23} & 0 & 0 & 0 & \xi_{9}
    \end{array}\right)
\end{equation}
Here the definitions of $\alpha$, $\beta$, $\eta$ and $\xi$ are the same as those in Eqs.~\eqref{eq:hamiltonian matrix}) and \eqref{eq:hamiltonian matrix-2}.
The environmental terms are now disentangled with resonant frequencies.
Due to the fact that all coherence terms (e.g. $\rho_{ij}=0$, $i\neq j$) would be zero, the evolution equation can also be simplified as 
\begin{equation}
    \frac{d}{dt}\left(\begin{array}{c}
        \rho_{11}(t)\\
        \rho_{22}(t)\\
        \rho_{33}(t)
    \end{array}\right)=\left(\begin{array}{ccc}
        -\gamma_{12} & \gamma_{12} & 0\\
        \gamma_{12} & -(\gamma_{12}+\gamma_{23}) & \gamma_{23}\\
        0 & \gamma_{23} & -\gamma_{23}
    \end{array}\right)\left(\begin{array}{c}
        \rho_{11}(t)\\
        \rho_{22}(t)\\
        \rho_{33}(t)
    \end{array}\right),
    \label{eq:three-level-no-electric}
\end{equation}
with 
\begin{equation}
    \gamma_{ij}=\eta\alpha_{ij}.
\end{equation}

Its analytical solution is 
\begin{equation}
    \begin{cases}
        \rho_{11}(t)=\frac{1}{3}-(\frac{\gamma_{12}+\gamma_{23}}{6\varOmega}-\frac{1}{6})\frac{(-\gamma_{12}+\gamma_{23}-\varOmega)e^{(-\gamma_{12}\text{-}\gamma_{23}-\varOmega)t}}{\gamma_{23}}\\
        -(\frac{-\gamma_{12}-\gamma_{23}}{6\varOmega}-\frac{1}{6})\frac{(-\gamma_{12}+\gamma_{23}+\varOmega)e^{(-\gamma_{12}-\gamma_{23}+\varOmega)t}}{\gamma_{23}}\\
        \rho_{22}(t)=\frac{1}{3}-(\frac{\gamma_{12}+\gamma_{23}}{6\varOmega}-\frac{1}{6})\frac{(\gamma_{12}+\varOmega)e^{(-\gamma_{12}-\gamma_{23}-\varOmega)t}}{\gamma_{23}}\\
        -(\frac{-\gamma_{12}-\gamma_{23}}{6\varOmega}-\frac{1}{6})\frac{(\gamma_{12}-\varOmega)e^{(-\gamma_{12}-\gamma_{23}+\varOmega)t}}{\gamma_{23}}\\
        \rho_{33}(t)=\frac{1}{3}+(\frac{\gamma_{12}+\gamma_{23}}{6\varOmega}-\frac{1}{6})e^{(-\gamma_{12}-\gamma_{23}-\varOmega)t}\\
        +(\frac{-\gamma_{12}-\gamma_{23}}{6\varOmega}-\frac{1}{6})e^{(-\gamma_{12}-\gamma_{23}+\varOmega)t}
    \end{cases},
\end{equation}
with $\varOmega=\sqrt{\gamma_{12}^{2}-\gamma_{12}\gamma_{23}+\gamma_{23}^{2}}$.

Then the following mathematical transformation is applied to eliminate non-zero diagonal elements
\begin{equation}
    \begin{cases}
        \widetilde{\rho_{11}}(t) & =\rho_{11}(t)-\frac{1}{3}\\
        \widetilde{\rho_{22}}(t) & =\rho_{22}(t)-\frac{1}{3}\\
        \widetilde{\rho_{33}}(t) & =\rho_{33}(t)-\frac{1}{3}
    \end{cases},\label{eq:transfers equation}
\end{equation}
leading to the new dynamic equation in the transformed space
\begin{equation}
    \frac{d}{dt}\left(\begin{array}{c}
        \widetilde{\rho_{11}}(t)\\
        \widetilde{\rho_{22}}(t)\\
        \widetilde{\rho_{33}}(t)
    \end{array}\right)=-\left(\begin{array}{ccc}
        0 & 2\gamma_{12} & \gamma_{12}\\
        2\gamma_{12}+\gamma_{23} & 0 & \gamma_{12}+2\gamma_{23}\\
        \gamma_{23} & 2\gamma_{23} & 0
    \end{array}\right)\left(\begin{array}{c}
        \widetilde{\rho_{11}}(t)\\
        \widetilde{\rho_{22}}(t)\\
        \widetilde{\rho_{33}}(t)
    \end{array}\right).
\end{equation}

Finally the original off-diagonal encoding schemes can be applied to extract pathway amplitudes,
which agrees well with those obtained by Dyson expansion (not shown here). 

\subsection{The three-level open quantum system with control field}

When there is control field, the necessary transformation is much more complicated.
The first step is to employ rotating-wave approximation (RWA)~\cite{RWA}.
The Hamiltonian in Eq.~\eqref{eq:hamiltonian matrix-2-1} in the interaction picture becomes
\begin{equation}
    \mathcal{H}_{I}=\left[\begin{array}{ccccccccc}
            -i\gamma_{12} & \varepsilon_{12} & 0 & -\varepsilon_{12} & i\gamma_{12} & 0 & 0 & 0 & 0\\
            \varepsilon_{12} & -id_{1} & \varepsilon_{23} & 0 & -\varepsilon_{12} & 0 & 0 & 0 & 0\\
            0 & \varepsilon_{23} & -\frac{i}{2}d_{2} & 0 & 0 & -\varepsilon_{12} & 0 & 0 & 0\\
            -\varepsilon_{12} & 0 & 0 & -id_{1} & \varepsilon_{12} & 0 & -\varepsilon_{23} & 0 & 0\\
            i\gamma_{12} & -\varepsilon_{12} & 0 & \varepsilon_{12} & -id_{2} & \varepsilon_{23} & 0 & -\varepsilon_{23} & i\gamma_{23}\\
            0 & 0 & -\varepsilon_{12} & 0 & \varepsilon_{23} & -id_{3} & 0 & 0 & -\varepsilon_{23}\\
            0 & 0 & 0 & -\varepsilon_{23} & 0 & 0 & -\frac{i}{2}d_{2} & \varepsilon_{12} & 0\\
            0 & 0 & 0 & 0 & -\varepsilon_{23} & 0 & \varepsilon_{12} & -id_{3} & \varepsilon_{23}\\
            0 & 0 & 0 & 0 & i\gamma_{23} & -\varepsilon_{23} & 0 & \varepsilon_{23} & -i\gamma_{23}
    \end{array}\right],
\end{equation}
with $d_{1}=(\gamma_{12}+\frac{\gamma_{23}}{2})$, $d_{2}=(\gamma_{12}+\gamma_{23})$,
$d_{3}=(\frac{\gamma_{12}}{2}+\gamma_{23})$, and $\varepsilon_{ij}$ being Rabi frequency.
In our simulations, the control field is taken to be of a Gaussian envelope
\begin{equation}
    \begin{aligned}
        E(t) & =e^{-\frac{(t-T/2)^{2}}{2\sigma^{2}}}A_{1}cos(\omega_{2}t+\varphi_{1})+e^{-\frac{(t-T/2)^{2}}{2\sigma^{2}}}A_{2}cos((\omega_{3}-\omega_{2})t+\varphi_{2})\\
        & =\frac{A_{1}}{2}e^{-\frac{(t-T/2)^{2}}{2\sigma^{2}}}(e^{i(\omega_{2}t+\varphi_{1})}+e^{-i(\omega_{2}t+\varphi_{1})})+\frac{A_{2}}{2}e^{-\frac{(t-T/2)^{2}}{2\sigma^{2}}}(e^{i((\omega_{3}-\omega_{2})t+\varphi_{2})}+e^{-i((\omega_{3}-\omega_{2})t+\varphi_{2})})
    \end{aligned},
\end{equation} 
leading to the Rabi frequencies $\varepsilon_{12}(t)=\frac{\mu_{12}A_{1}}{2}e^{-\frac{(t-T/2)^{2}}{2\sigma^{2}}}$
and $\varepsilon_{23}(t)=\frac{\mu_{23}A_{2}}{2}e^{-\frac{(t-T/2)^{2}}{2\sigma^{2}}}$. 

To eliminate the non-zero diagonal elements, the following transformation is adopted 
\begin{equation}
    \begin{cases}
        \widetilde{\rho_{11}}(t) & =\rho_{11}(t)-\frac{1}{3}\\
        \widetilde{\rho_{22}}(t) & =\rho_{22}(t)-\frac{1}{3}\\
        \widetilde{\rho_{33}}(t) & =\rho_{33}(t)-\frac{1}{3}\\
        \widetilde{\rho_{12}}(t) & =\rho_{12}(t)e^{(\gamma_{12}+\frac{\gamma_{23}}{2})t}\\
        \widetilde{\rho_{13}}(t) & =\rho_{13}(t)e^{\frac{\gamma_{12}+\gamma_{23}}{2}t}\\
        \widetilde{\rho_{21}}(t) & =\rho_{21}(t)e^{(\gamma_{12}+\frac{\gamma_{23}}{2})t}\\
        \widetilde{\rho_{23}}(t) & =\rho_{23}(t)e^{(\frac{\gamma_{12}}{2}+\gamma_{23})t}\\
        \widetilde{\rho_{31}}(t) & =\rho_{31}(t)e^{\frac{\gamma_{12}+\gamma_{23}}{2}t}\\
        \widetilde{\rho_{32}}(t) & =\rho_{32}(t)e^{(\frac{\gamma_{12}}{2}+\gamma_{23})t}
    \end{cases}.\label{eq:transfers equation-1}
\end{equation}
leading to the new Hamiltonian
\begin{align}
    \widetilde{\mathcal{H}_{I}}(t) & =-i\left(\begin{array}{ccccccccc}
        0 & -i\mathcal{L}(-t) & 0 & i\mathcal{L}(-t) & 2\gamma_{12} & 0 & 0 & 0 & \gamma_{12}\\
        -i\mathcal{L}(t) & 0 & -i\mathcal{O}_{1}(t) & 0 & i\mathcal{L}(t) & 0 & 0 & 0 & 0\\
        0 & -i\mathcal{O}_{1}(-t) & 0 & 0 & 0 & i\mathcal{O}_{2}(-t) & 0 & 0 & 0\\
        i\mathcal{L}(t) & 0 & 0 & 0 & -i\mathcal{L}(t) & 0 & i\mathcal{O}_{1}(t) & 0 & 0\\
        2d_{1} & i\mathcal{L}(-t) & 0 & -i\mathcal{L}(-t) & 0 & -i\mathcal{F}(-t) & 0 & i\mathcal{F}(-t) & 2d_{3}\\
        0 & 0 & i\mathcal{O}_{2}(t) & 0 & -i\mathcal{F}(t) & 0 & 0 & 0 & i\mathcal{F}(t)\\
        0 & 0 & 0 & i\mathcal{O}_{1}(-t) & 0 & 0 & 0 & -i\mathcal{O}_{2}(-t) & 0\\
        0 & 0 & 0 & 0 & i\mathcal{F}(t) & 0 & -i\mathcal{O}_{2}(t) & 0 & -i\mathcal{F}(t)\\
        \gamma_{23} & 0 & 0 & 0 & 2\gamma_{23} & i\mathcal{F}(-t) & 0 & -i\mathcal{F}(-t) & 0
    \end{array}\right),
    \label{eq:three-electirc-hamiltonian}
\end{align}
which governs the evolution equation 
\begin{equation}
    i\frac{d}{dt}\left(\begin{array}{c}
        \widetilde{\rho_{11}}(t)\\
        \widetilde{\rho_{12}}(t)\\
        \widetilde{\rho_{13}}(t)\\
        \widetilde{\rho_{21}}(t)\\
        \widetilde{\rho_{22}}(t)\\
        \widetilde{\rho_{23}}(t)\\
        \widetilde{\rho_{31}}(t)\\
        \widetilde{\rho_{32}}(t)\\
        \widetilde{\rho_{33}}(t)
    \end{array}\right)=\widetilde{\mathcal{H}_{I}}\left(\begin{array}{c}
        \widetilde{\rho_{11}}(t)\\
        \widetilde{\rho_{12}}(t)\\
        \widetilde{\rho_{13}}(t)\\
        \widetilde{\rho_{21}}(t)\\
        \widetilde{\rho_{22}}(t)\\
        \widetilde{\rho_{23}}(t)\\
        \widetilde{\rho_{31}}(t)\\
        \widetilde{\rho_{32}}(t)\\
        \widetilde{\rho_{33}}(t)
    \end{array}\right).
\end{equation}

Here $\mathcal{L}(t)=\varepsilon_{12}e^{(\gamma_{12}+\frac{\gamma_{23}}{2})t}$,
$\mathcal{F}(t)=\varepsilon_{23}e^{(\frac{\gamma_{12}}{2}+\gamma_{23})t}$,
$\mathcal{O}_{1}(t)=\varepsilon_{23}e^{\frac{\gamma_{12}}{2}t}$ and
$\mathcal{O}_{2}(t)=\varepsilon_{12}e^{\frac{\gamma_{23}}{2}t}$.

\begin{figure}
    \includegraphics[width=0.8\columnwidth]{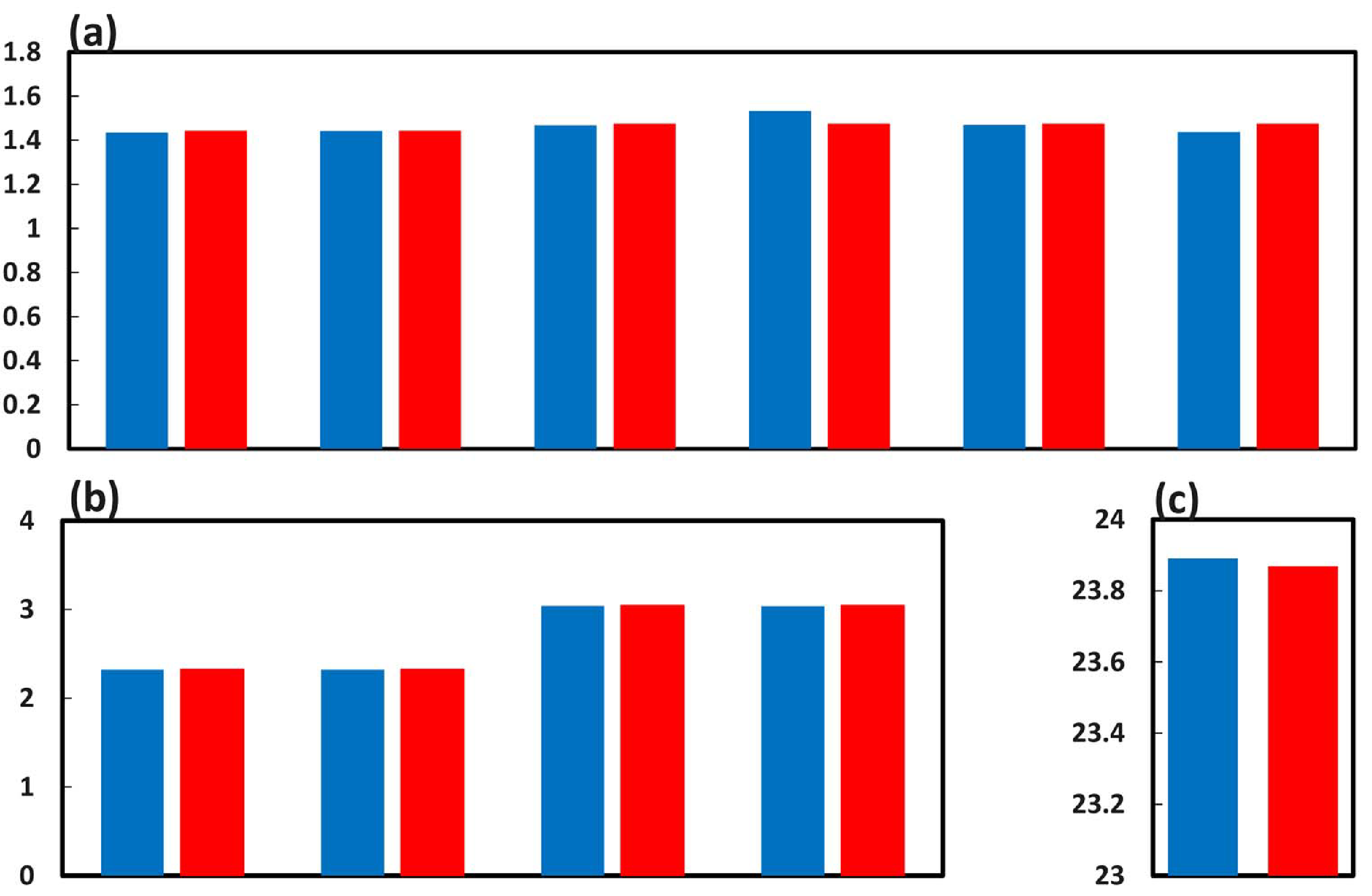}
    \caption{\label{fig:new-three-level-system}
    (Color online) Magnitudes of significant pathways with different methods.
    The vertical axis denotes the magnitude, with the values in panel 
    (a) scaled by a factor of 1000 and those in panels
    (b) and (c) scaled by a factor of 100.
    Panels (a), (b) and (c) are, respectively, the dipole-induced pathways, the dipole-environment-induced pathways and the exclusively environment-induced pathways.
    The pathway indices along the horizontal axis are arranged in the same order as in Tab.~\ref{tab:New-matrix-three-level}.
    The blue and red bars are, respectively, obtained by the improved HE-OD method and Dyson expansion.}
\end{figure}

Then methods of HE-OD with the normal off-diagonal encoding scheme
and Dyson expansion can be employed to obtain the pathway amplitudes.
The results are shown in Tab.~\ref{tab:New-matrix-three-level} and Fig.~\ref{fig:new-three-level-system}. 
\begin{table}
    \caption{\label{tab:New-matrix-three-level}
    Magnitudes and phases of significant quantum pathways for the three-level open system when both dissipation and control fields are present.
    The values in the parentheses are the phases.}

\begin{tabular}{cc>{\centering}p{3cm}c}
    \hline 
    Type & Pathways & Amplitude

    (Phase) & Integration\tabularnewline
    \hline 
    \multirow{6}{*}{dipole } & $\left.\left|\widetilde{11}\right\rangle \right\rangle \rightarrow\left.\left|\widetilde{12}\right\rangle \right\rangle \rightarrow\left.\left|\widetilde{13}\right\rangle \right\rangle \rightarrow\left.\left|\widetilde{23}\right\rangle \right\rangle \rightarrow\left.\left|\widetilde{33}\right\rangle \right\rangle $ & $1.435\times10^{-3}$

    ($4.926\times10^{-14}$) & $1.443\times10^{-3}$\tabularnewline
    & $\left.\left|\widetilde{11}\right\rangle \right\rangle \rightarrow\left.\left|\widetilde{21}\right\rangle \right\rangle \rightarrow\left.\left|\widetilde{31}\right\rangle \right\rangle \rightarrow\left.\left|\widetilde{32}\right\rangle \right\rangle \rightarrow\left.\left|\widetilde{33}\right\rangle \right\rangle $ & $1.442\times10^{-3}$

    ($4.707\times10^{-13}$) & $1.443\times10^{-3}$\tabularnewline
    & $\left.\left|\widetilde{11}\right\rangle \right\rangle \rightarrow\left.\left|\widetilde{12}\right\rangle \right\rangle \rightarrow\left.\left|\widetilde{22}\right\rangle \right\rangle \rightarrow\left.\left|\widetilde{32}\right\rangle \right\rangle \rightarrow\left.\left|\widetilde{33}\right\rangle \right\rangle $ & $1.468\times10^{-3}$

    ($3.732\times10^{-14}$) & $1.476\times10^{-3}$\tabularnewline
    & $\left.\left|\widetilde{11}\right\rangle \right\rangle \rightarrow\left.\left|\widetilde{21}\right\rangle \right\rangle \rightarrow\left.\left|\widetilde{22}\right\rangle \right\rangle \rightarrow\left.\left|\widetilde{23}\right\rangle \right\rangle \rightarrow\left.\left|\widetilde{33}\right\rangle \right\rangle $ & $1.533\times10^{-3}$

    ($5.196\times10^{-13}$) & $1.476\times10^{-3}$\tabularnewline
    & $\left.\left|\widetilde{11}\right\rangle \right\rangle \rightarrow\left.\left|\widetilde{12}\right\rangle \right\rangle \rightarrow\left.\left|\widetilde{22}\right\rangle \right\rangle \rightarrow\left.\left|\widetilde{23}\right\rangle \right\rangle \rightarrow\left.\left|\widetilde{33}\right\rangle \right\rangle $ & $1.470\times10^{-3}$

    ($-3.164\times10^{-13}$) & $1.476\times10^{-3}$\tabularnewline
    & $\left.\left|\widetilde{11}\right\rangle \right\rangle \rightarrow\left.\left|\widetilde{21}\right\rangle \right\rangle \rightarrow\left.\left|\widetilde{22}\right\rangle \right\rangle \rightarrow\left.\left|\widetilde{32}\right\rangle \right\rangle \rightarrow\left.\left|\widetilde{33}\right\rangle \right\rangle $ & $1.438\times10^{-3}$

    ($-3.581\times10^{-13}$) & $1.476\times10^{-3}$\tabularnewline
    \hline 
    \multirow{4}{*}{dipole-dissipation} & $\left.\left|\widetilde{11}\right\rangle \right\rangle \rightarrow\left.\left|\widetilde{12}\right\rangle \right\rangle \rightarrow\left.\left|\widetilde{22}\right\rangle \right\rangle \rightarrow\left.\left|\widetilde{33}\right\rangle \right\rangle $ & $2.322\times10^{-2}$

    ($1.097\times10^{-14}$) & $2.326\times10^{-2}$\tabularnewline
    & $\left.\left|\widetilde{11}\right\rangle \right\rangle \rightarrow\left.\left|\widetilde{21}\right\rangle \right\rangle \rightarrow\left.\left|\widetilde{22}\right\rangle \right\rangle \rightarrow\left.\left|\widetilde{33}\right\rangle \right\rangle $ & $2.322\times10^{-2}$

    ($6.223\times10^{-14}$) & $2.326\times10^{-2}$\tabularnewline
    & $\left.\left|\widetilde{11}\right\rangle \right\rangle \rightarrow\left.\left|\widetilde{22}\right\rangle \right\rangle \rightarrow\left.\left|\widetilde{23}\right\rangle \right\rangle \rightarrow\left.\left|\widetilde{33}\right\rangle \right\rangle $ & $3.042\times10^{-2}$

    ($1.382\times10^{-14}$) & $3.045\times10^{-2}$\tabularnewline
    & $\left.\left|\widetilde{11}\right\rangle \right\rangle \rightarrow\left.\left|\widetilde{22}\right\rangle \right\rangle \rightarrow\left.\left|\widetilde{32}\right\rangle \right\rangle \rightarrow\left.\left|\widetilde{33}\right\rangle \right\rangle $ & $3.040\times10^{-2}$

    ($2.191\times10^{-14}$) & $3.045\times10^{-2}$\tabularnewline
    \hline 
    dissipation & $\left.\left|\widetilde{11}\right\rangle \right\rangle \rightarrow\left.\left|\widetilde{22}\right\rangle \right\rangle \rightarrow\left.\left|\widetilde{33}\right\rangle \right\rangle $ & $2.389\times10^{-1}$

    ($6.125\times10^{-15}$) & $2.3868\times10^{-1}$\tabularnewline
    \hline 
\end{tabular}
\end{table}

\subsection{Generalization of the transformation and comparison of pathways in the two spaces}

The above transformation can be generalized to an N-level open quantum system.
Its equilibrium state in the pure dissipation case can be described by $\rho_{mm}^{eq}=c_{m}$ and $\rho_{mn}^{eq}=0\,\left(m\neq n,\,1\leq m,n\leq N\right)$, and the transformation without control field is $\widetilde{\rho_{mm}}(t)=\rho_{mm}(t)-c_{m}$.
When the control field is present, the transformation becomes to $\widetilde{\rho_{mm}}(t)=\rho_{mm}(t)-c_{m}$
and $\widetilde{\rho_{mn}}(t)=\rho_{mn}(t)e^{i(\mathcal{H_{I}})_{mn,mn}t}\,\,\left(m\neq n,\,1\leq m,n\leq N\right)$,
where $\mathcal{H}_{I}$ is the Hamiltonian in the Liouville space in the interaction picture. 

The definition of pathways in the transformed space is clearer than that in the original Liouville space.
The inconsistency between the amplitudes extracted by HE-OD and the integration from Dyson Expansion is also avoided.
However, it is still quite interesting to compare the pathways extracted by HE-OD in the two spaces.
The control mechanism can be described by the transition pathways linking the initial and target states.
With the original and transformed spaces, the states are, respectively, described in the two bases of $\left.\left|mn\right\rangle \right\rangle $ and $\left.\left|\widetilde{m}\widetilde{n}\right\rangle \right\rangle \,\,\left(1\leq m,n\leq N\right)$.
The two methods both have advantages and disadvantages.
The initial state is simpler with the original HE-OD method in the original space,
but diagonal encodings are necessary due to non-zero diagonal elements in the Hamiltonian,
which will lead to self-to-self transitions.
The improved HE-OD method may lead to more complex pathways due to the different description of the initial state in the transformed space.
For the above three-level open system, the initial state is $\left.\left|11\right\rangle \right\rangle $
with the original method, while is $\frac{2}{3}\left.\left|\widetilde{1}\widetilde{1}\right\rangle \right\rangle -\frac{1}{3}\left.\left|\widetilde{2}\widetilde{2}\right\rangle \right\rangle -\frac{1}{3}\left.\left|\widetilde{3}\widetilde{3}\right\rangle \right\rangle $
with the improved method. Thus with the new method, pathways of the types $\left.\left|\widetilde{2}\widetilde{2}\right\rangle \right\rangle \rightarrow\ldots\rightarrow\left|\widetilde{3}\widetilde{3}\right\rangle $
and $\left.\left|\widetilde{3}\widetilde{3}\right\rangle \right\rangle \rightarrow\ldots\rightarrow\left|\widetilde{3}\widetilde{3}\right\rangle $
will appear in the mechanism analysis in addition to those of the type $\left.\left|\widetilde{1}\widetilde{1}\right\rangle \right\rangle \rightarrow\ldots\rightarrow\left|\widetilde{3}\widetilde{3}\right\rangle $,
while only one type (\emph{i.e.}$\left.\left|11\right\rangle \right\rangle \rightarrow\ldots\rightarrow\left|33\right\rangle $)
is present with the original method. The main advantage of the improved method in this paper is that it avoids self-to-self transitions and thus can obtain the correct pathway amplitudes without diagonal encodings and avoid the inconsistency with Dyson expansion. 

\section{Conclusions and discussions\label{sec:Conclusions-and-discussions}}

HE-OD can extract pathway amplitudes and is an efficient way to investigate mechanisms in the control of quantum dynamics.
For closed systems, the original off-diagonal encodings in HE-OD can induce the same pathway amplitudes as those by Dyson expansion. When moving to open quantum systems, some discrepancy appears.
The underlying reason is related to the fact that the environmental terms are present in the diagonal Hamiltonian elements.
Off-diagonal encodings can not differentiate transitions induced by these elements. The discrepancy can be diminished by diagonal encodings.
However, this encoding scheme will introduce self-to-self transitions, which are counter-intuitive and inconvenient to be described mathematically.
In this work, an improved HE-OD methodology is proposed to avoid such transitions in the resulted pathways.
Firstly, the original Hamiltonian is transformed into the interaction picture to disentangle the environmental terms and resonant frequencies;
Then a proper mathematical transformation is performed to eliminate the diagonal elements;
Finally, off-diagonal encodings can be employed to differentiate pathways expressed in the new basis.
The improved HE-OD method shows good consistence with the accurate Dyson expansion.
A three-level open quantum system is taken as example to illustrate the effectiveness of our proposed methodology,
which can be generalized to multi-level open quantum systems. 
\begin{acknowledgments}
    The authors acknowledge support by the National Natural Science Foundation of China 
    (Grants No. 61720106009, No. 61773359, No. 51401121, No. 61403362, 61374091, and No. 61473199).
    F. Shuang thanks the Leader talent plan of the Universities in Anhui Province and the CAS Interdisciplinary Innovation Team of the Chinese Academy of Sciences for financial support.
\end{acknowledgments}


\begin{thebibliography}{10}
    \bibitem{key-1-1}H. Rabitz, R. de Vivie-Riedle, M. Motzkus, and K.
        Kompa, Science 288, 824 (2000). 

    \bibitem{key-1-2}J. L. Herek, W. Wohlleben, R. J. Cogdell, D. Zeidler,
        and M. Motzkus, Nature 417, 533 (2002). 

    \bibitem{key-1-3}Y. Silberberg, Annu. Rev. Phys. Chem. 60, 277 (2009). 

    \bibitem{key-1-4}J. Zhang, Y. X. Liu, R. B. Wu, C. W. Li, and T.
        J. Tarn, Phys. Rev. A 82, 022101 (2010). 

    \bibitem{key-1-5}J. Zhang, Y. X. Liu, and F. Nori, Phys. Rev. A 79,
        052102 (2009). 

    \bibitem{key-1-6}H. M. Wiseman and A. C. Doherty, Phys. Rev. Lett.
        94, 070405 (2005). 

    \bibitem{key-1-7}H. Mabuchi, Phys. Rev. A 78, 032323 (2008). 

    \bibitem{key-1-8}G. G. Gillett, R. B. Dalton, B. P. Lanyon, M. P.
        Almeida, M. Barbieri, G. J. Pryde, J. L. O\textquoteright{} Brien,
        K. J. Resch, S. D. Bartlett, and A. G. White, Phys. Rev. Lett. 104,
        080503 (2010). 

    \bibitem{key-1-9}C. Altafini and F. Ticozzi, IEEE Trans. Auto. Cont.
        57, 1898 (2012). 

    \bibitem{key-1-10}C. Brif, M. D. Grace, M. Sarovar, and K. C. Young,
        New J. Phys. 16, 065013 (2014). 

    \bibitem{key-1-11}S. J. Glaser, U. Boscain, T. Calarco, C. P. Koch,
        Walter Kockenberger, R. Kosloff, I. Kuprov, B. Luy, S. Schirmer, T.
        Schulte-Herbruggen, D. Sugny, and F. K. Wilhelm, Eur. Phys. J. D 69,
        279 (2015).

    \bibitem{m2}J. Petersen, R. Mitri$\acute{c}$, V. Bona$\check{c}$i$\acute{c}$-Kouteck$\acute{y}$,
        J. P. Wolf, J. Roslund, H. Rabitz, Phys. Rev. Lett. 105, 073003 (2010)

    \bibitem{m3}T. Brixner, N. H. Damrauer, P. Niklaus, G. Gerber, Nature,
        414, 57 (2001)

    \bibitem{m4}R. J. Gordon, S. A. Rice, Annu. Rev. Phys. Chem. 48,
        601 (1997)

    \bibitem{m5}D. G. Kuroda, C. P. Singh, Z. Peng, V. D. Kleiman, Science
        326, 263 (2009)

    \bibitem{m6}K. Hoki, P. Brumer, Phys. Rev. Lett. 95, 168305 (2005)

    \bibitem{m7}P. van der Walle, M. T. W. Milder, L. Kuipers, J. L.
        Herek, Proc. Natl. Acad. Sci. USA 106, 7714 (2009)

    \bibitem{key-1}C. Daniel, J. Full, L. Gonz$\acute{a}$lez, C. Lupulescu,
        J. Manz, A. Merli, $\check{S}$. Vajda, L. W$\ddot{o}$ste, Science
        299, 536 (2003)

    \bibitem{key-2}M. Delor, T. Keane, P. A. Scattergood, I. V. Sazanovich,
        G. M. Greetham, M. Towrie, A. J. H. M. Meijer, J. A. Weinstein, Nature
        Chem. 7, 689 (2015)

    \bibitem{key-3}X. Xing, R. Rey-de-Castro, H. Rabitz, New J. Phys.
        16, 125004 (2014)

    \bibitem{PRA2014}F. Gao, R. Rey-de-Castro, A. M. Donovan, J. Xu,
        Y. Wang, H. Rabitz, F. Shuang, Phys. Rev. A 89, 023416 (2014)

    \bibitem{key-1-14shuang-gao}F. Gao, Y. Wang, R. Rey-de-Castro, H.
        Rabitz, and F. Shuang, Phys. Rev. A 92, 033423 (2015).

    \bibitem{key-1-12}A. Mitra and H. Rabitz, Phys. Rev. A 67, 033407
        (2003).

    \bibitem{key-1-13}R. Rey-de-Castro, Z. Leghtas, and H. Rabitz, Phys.
        Rev. Lett. 110, 223601 (2013). 

    \bibitem{RobertoNJP}R. Rey-de-Castro, R. Cabrera, D. I. Bonder, and
        H. Rabitz, New J. Phys. 15, 025032 (2013).

    \bibitem{key-1-15gao}F. Gao, R. Rey-de-Castro, Y. Wang, H. Rabitz,
        and F. Shuang.Phys. Rev.A 93,053407 (2016)

    \bibitem{RWA}I. I. Rabi, N. F. Ramsey, and J. Schwinger, Rev. Mod.
        Phys. 26, 167 (1954).
\end{thebibliography}
\end{document}